\documentclass[graphicx,aps,pre,epsfig,article,showpacs]{revtex4}
\usepackage{graphicx}
\usepackage{amsmath}
\usepackage{appendix}

\begin{document}

\title{A Stochastic Local Search algorithm for distance-based
phylogeny reconstruction}

\author{$^1$F. Tria, $^2$E. Caglioti, $^{1,3}$V. Loreto and $^1$A. Pagnani}

\affiliation{$^1$ ISI Foundation, Viale Settimio Severo 65, Villa
  Gualino, I-10133 Torino, Italy, \\$^2$Dipartimento di Matematica,
  Sapienza Universit\`a di Roma, Piazzale Aldo Moro 5, 00185 Roma,
  Italy, \\$^3$Dipartimento di Fisica, Sapienza Universit\`a di Roma,
  Piazzale Aldo Moro 5, 00185 Roma, Italy}

\pacs{89.75.Hc, 87.23.Kg, 05.10.Ln, 02.70.Uu}

\begin{abstract}
  In many interesting cases the reconstruction of a correct phylogeny
  is blurred by high mutation rates and/or horizontal transfer events.
  As a consequence a divergence arises between the true evolutionary
  distances and the differences between pairs of taxa as inferred from
  available data, making the phylogenetic reconstruction a challenging
  problem. Mathematically this divergence translates in a loss of
  additivity of the actual distances between taxa. In distance-based
  reconstruction methods, two properties of additive distances were
  extensively exploited as antagonist criteria to drive phylogeny
  reconstruction: on the one hand a local property of quartets, i.e.,
  sets of four taxa in a tree, the four-points condition; on the other
  hand a recently proposed formula that allows to write the tree
  length as a function of the distances between taxa, the Pauplin's
  formula. Here we introduce a new reconstruction scheme, that
  exploits in a unified framework both the four-points condition and
  the Pauplin's formula. We propose, in particular, a new general
  class of distance-based Stochastic Local Search algorithms, which
  reduces in a limit case to the minimization of the Pauplin's length.
  When tested on artificially generated phylogenies our Stochastic
  Big-Quartet Swapping algorithmic scheme significantly outperforms
  state-of-art distance-based algorithms in cases of deviation from
  additivity due to high rate of back mutations. A significant
  improvement is also observed with respect to the state-of-art
  algorithms in case of high rate of horizontal transfer. 
\end{abstract}

\maketitle




\section{Introduction}

Phylogenetic methods have recently been rediscovered in several
interesting areas among which immunodynamics, epidemiology and many
branches of evolutionary dynamics. The reconstruction of phylogenetic
trees belongs to a general class of inverse problems whose relevance
is now well established in many different disciplines ranging from
biology to linguistics and social
sciences~\cite{Gray2003,css_2009,pybus_2009,warnow_science_2009}. In
a generic inverse problem one is given with a set of data and has to
infer the most likely dynamical evolution process that presumably
produced the given data set. The relevance of inverse problems has
been certainly triggered by the fast progress in data-revealing
technologies. In molecular biology, for instance, a great amount of
genomes data are available thanks to the new high-throughput methods
for genome analysis~\cite{ragousis_2009}. In historical
linguistics~\cite{renfrew_2000} a remarkable effort has been recently
done for the compilation of corpora of homologous features (lexical,
phonological, syntactic) or characters for many different languages.

Although phylogenetic reconstruction is not a novel topic, dealing
with not purely tree-like processes and identifying the possible
sources of non-additivity and their effects in a given dataset is
still an open and challenging
problem~\cite{felsenstein2004,GASCUEL_2007}.

Here we focus on distance-based
methods~\cite{cavalli_edwards67,fitch_margoliash67} and investigate
how deviations from additivity affect their performances. In
distance-based methods only distances between leaves are considered,
and all the information possibly encoded in the combinatorial
structure of the character states is lost. Despite their simplicity,
distance-based methods are still widely used thanks to their
computational efficiency, but a solid theoretical understanding on the
limitation of their applicability is still lacking. One of the most
popular distance-based reconstruction algorithm, Neighbor-Joining
(NJ)~\cite{NJ}, was proposed in the late {\em 80's}, but it is only
recently that its theoretical background was put on a more solid
basis~\cite{Atteson97, Gascuel2006, Mihaescu2007}. Another step
toward a better understanding of distance-based methods was obtained
thanks to an interesting property of additive distances, the Pauplin's
formula~\cite{pauplin2000}. This property has been used in the
formulation of a novel algorithmic strategy with improved performances
(FastME)~\cite{fastme_2002}. In parallel, another fundamental
property of additive trees, the {\em four-points}
condition~\cite{Gusfield97}, has been extensively exploited in
distance-based phylogenetic reconstruction
methods~\cite{erdos_1998,snir_2008}. Both the Pauplin's formula and
the {\em four-points} condition will be discussed in details below.

Here we propose a new approach that combines the {\em four-points}
condition and the Pauplin's formula in a Stochastic Local Search (SLS)
scheme that we name Stochastic Big-Quartet Swapping (SBiX) algorithm.
SLS~\cite{booksls} algorithms transverse the search space of a given
problem in a systematic way, allowing for a sampling of low cost
configurations. SLS algorithms start from a randomly chosen initial
condition. Subsequently the elementary step is a move from a
configuration to a neighboring one. Each move is determined by a
decision based on local knowledge only. Typically the decision is
taken combining, with a given a priori probability, a greedy step
(i.e., a step that reduces the local cost contribution), with a random
one where the local cost is not taken into account. SLS algorithms
have been widely used in solving complex combinatorial optimization
problems such as Satisfiability, Coloring, MAX-SAT, and Traveling
Salesman Problem~\cite{booksls}.

At the heart of our new algorithmic scheme (named SBiX) there is the
notion of {\em quartet frustration}, a quantitative measure of how
good a given configuration is, in the space of trees. Following a
concept already introduced in~\cite{snir_2008}, we weight the
different quartets according to their length, in order to reduce the
effect of those which are more likely to undergo double mutations. The
strength of our approach comes from a combination of this strategy
with a Pauplin's like one, weighting each quartet according to a
purely topological property.

We tested the performances of the proposed reconstruction algorithm,
i.e., the ability to reconstruct the true topology, in the presence of
high levels of deviation from additivity due to both horizontal
transfer and back-mutation processes. We use both a very simple model
to generate artificial phylogenies of binary sequences, and the more
realistic Kimura two-parameters model, considering sequences with
$q$-state sites, with $q=4$. We have evidence that the performance of
our algorithm does not rely on the particular evolutive process giving
rise to the phylogeny, nor on the particular representation of the
taxa. We find that when the lack of additivity arises from high
mutation rates (and consequently high probability of back mutations),
our algorithm significantly outperforms the state-of-art
distance-based algorithms. When the lack of additivity arises from
high rate of horizontal transfer events, our algorithm performs better
than the algorithms we considered as competing ones.

We show results both for a greedy and a simulated-annealing-like
strategy of our algorithm, the former being significantly faster than
the latter and with comparable performances. The SBiX algorithm has a
complexity of $O(N^4)$, which is higher that the one of the
distance-based algorithms used as competitors for comparing the
performances. Nevertheless, the prefactor of the greedy version is so
low that our algorithm is fast enough to reconstruct large phylogenies
e.g., of a few thousands taxa, in a time remarkably slower with
respect to any character based reconstruction algorithm. A comparison
of the running time and of the performances of our algorithm with a
popular character based one, MrBayes~\cite{MrBayes}, is reported
respectively in the Appendix II and in the section Results. We show
that, while the performances of the two algorithms are comparable, the
running time of MrBayes greatly exceed ours, becoming comparable when
the reconstruction becomes in practise unfeasible, that is for running
times of the order of some thousand of years, and number of taxa of
the order of one hundred thousand.

\section{Methods}
\label{sec:add}
\subsection{Additivity and the four-points condition}
A distance matrix is said to be additive if it can be constructed as
the sum of a tree's branches lengths.  Two fundamental and widely used
properties of additive distances are the Pauplin's
formula~\cite{pauplin2000} and the {\em four-points}
condition~\cite{Gusfield97}. For the sake of clarity, we recall them
here.

\begin{figure}
\centerline{\includegraphics[width=0.49\textwidth]{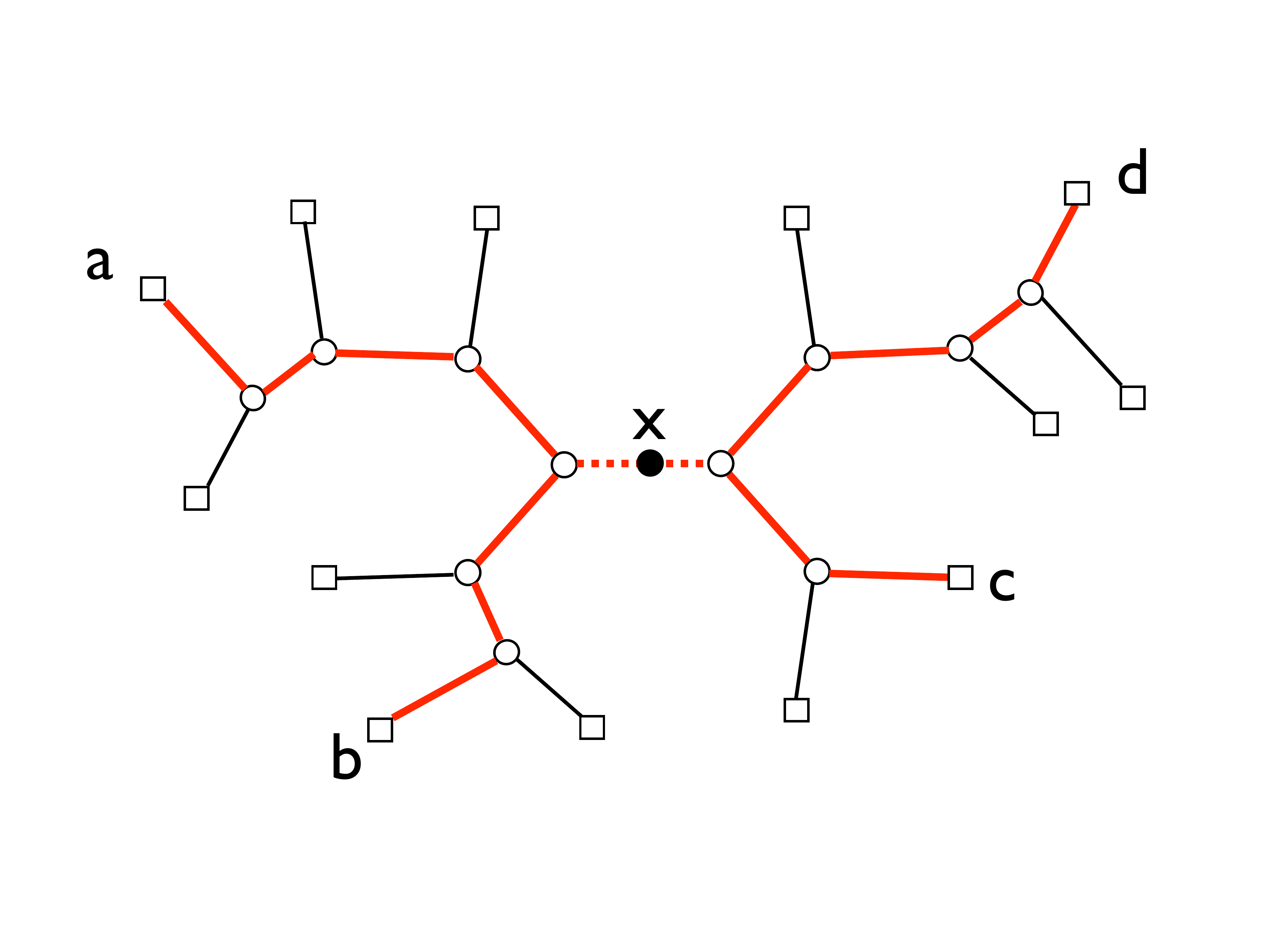}}
\caption{{\bf Quartet definition.} The quartet $a,b,c,d$ induces an
    internal edge $x$ that divides the tree in two parts. All paths
    joining any pair of sites sitting on opposite parts of the tree
    pass through $x$.}
\label{fig:tree}
\end{figure}

Given any quadruplet of taxa $a,b,c,d$, let $D_1=d(a,b)+d(c,d)$,
$D_2=d(a,c)+d(b,d)$, and $D_3=d(a,d)+d(b,c)$ be the three possible
pairs of distances between the four taxa. A matrix $M$ is additive if
and only if $D_1<D_2=D_3$ or $D_2<D_1=D_3$ or
$D_3<D_1=D_2$\footnote{It is important to remark here that the
  four-points condition is an equivalent definition of additivity.
  That is: a distance matrix is additive if and only if the
  four-points condition is satisfied.}. When considering experimental
data, additivity is almost always violated and so is the {\em
  four-points} condition. In order to set up a robust method for
phylogeny reconstruction based on the {\em four-points} condition, we
need to relax the notion of additivity and to quantify violations in a
suitable way.  For any four taxa $a,b,c,d$ such that $a,b$ are on one
side of the tree and $c,d$ on the other (as in fig.~\ref{fig:tree}),
the quartet $(ab:cd)$ is said to satisfy the {\em weak four-points}
condition if $D_1=\min(D_1,D_2,D_3)$ (where $D_1,D_2,D_3$ are defined
as above).  It is easy to prove that if the distance matrix $M$ is
additive a unique tree exists in which all quartets satisfy the {\em
  weak four-points} condition and this tree is the correct one. Many
algorithms have been proposed that exploit this {\em weak four-points}
condition, one of the most promising being, for instance, the
short-quartet method~\cite{erdos_1998,snir_2008}.

\subsection{Pauplin's distance}

Another remarkable property of an additive tree is the possibility to
compute its total length $L$, defined as the sum of all its branches
lengths, through a formula, due to Pauplin~\cite{pauplin2000}, that
only uses distances between taxa:
\begin{equation}
\label{eq:pauplin}
L_P= \sum_{a<b} 2^{-t(a,b)}M(a,b),
\end{equation}

\noindent where $t(a,b)$ is the number of nodes on the path connecting
$a$ and $b$, i.e., their {\em topological distance}. For additive
trees $L\equiv L_P$, but even when the {\em four-points} condition is
no longer satisfied, $L_P$ is a particularly good approximation for
the tree length~\cite{fastme_2002}. Furthermore, it is recognized
that for distance matrices sufficiently ``close'' to
additivity~\cite{Atteson97} the correct phylogeny minimizes
$L_P$~\cite{Mihaescu2007,Bordewich_2009}. This principle is used in an implicit way
in Neighbor-Joining~\cite{NJ} and more explicitly in
a new generation distance-based algorithm, FastME~\cite{fastme_2002}.
When departure from additivity is too strong $L_P$ is no longer a good
functional to minimize in order to recover the correct tree.

\subsection{Violations of additivity}

Violations of additivity can arise both from experimental noise and
from properties of the evolutionary process the data come from. We
here consider two of the main sources of violations of the latter type
that can either occur together or singularly. (i) {\em back-mutation}:
in particularly long phylogenies, here the time-scale being set by the
mutation rate, a single character can experience multiple mutations.
In this case the distances between taxa are no longer proportional to
their evolutionary distances; we will use in the following the
expression {\em back-mutation} as synonymous of multiple mutation on
the same site; (ii) {\em horizontal transfer}: the reconstruction of a
phylogeny from data underlies the assumption that information flows
{\em vertically} from ancestors to offsprings. However, in many
processes information flows also {\em horizontally}. Horizontal (or
lateral) gene transfers~\cite{treeoflifepnas2005}
are often well known
confounding factors for a correct phylogenetic inference.

\subsection{The Stochastic Big-Quartet Swapping algorithm
  \label{sec:algo}}

Here we describe the structure of our Stochastic Big-Quartet Swapping
(SBiX) algorithm.  As already mentioned this algorithm crucially
exploits both the {\em weak four-points} condition and the Pauplin's
distance, featuring a larger robustness with respect to violations of
additivity if compared to algorithms based separately on the {\em weak
  four-points} condition or on the minimization of the Pauplin's
length.

The general structure of the SBiX algorithm is as follows:

\begin{itemize}
\item[$1$] start with a tree topology for the given set of
  taxa;
\item[$2$] update the tree topology by local elementary rearrangements
  through sub-trees quartet swapping;
\item[$3$] repeat point $2$ till convergence is reached;
\end{itemize}

\begin{figure}
  \centerline{\includegraphics[width=0.49\textwidth]{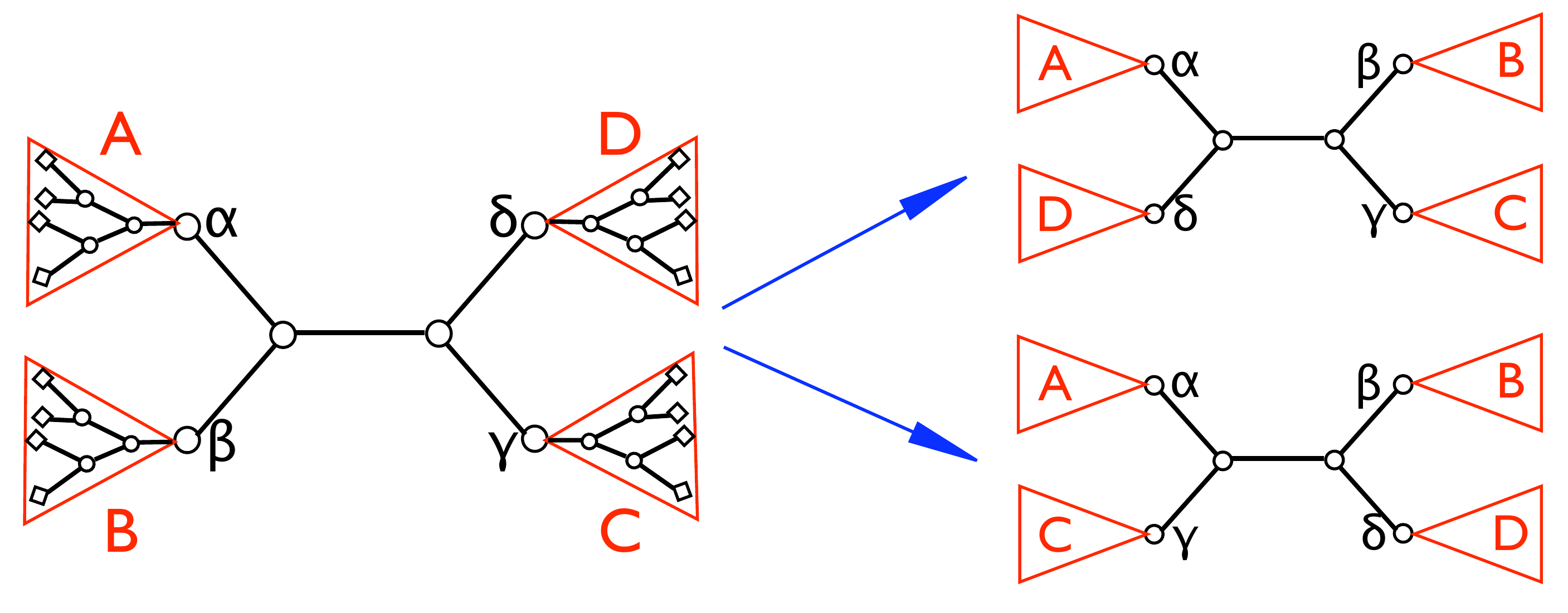}}
  \caption{{\bf Elementary move of our Stochastic Local Search scheme.} Any internal edge defines the four
    subtrees $A,B,C,D$ rooted respectively on
    $\alpha,\beta,\gamma,\delta$.  Here, the initial reference
    configuration $((A,B),(C,D))$ displayed on the left, leads to two
    possible rewiring: (i) swap the pair $B \leftrightarrow D$
    (right-upper panel), (ii) swap the pair $B \leftrightarrow C $
    (right-lower panel).}
  \label{fig:mcstep}
\end{figure}

\subsection{Sub-trees (big-)quartets swaps}

We analyze the three points in details  in the following:

\begin{itemize}
\item[$1$] In the simulated-annealing-like version of the algorithm we
  start with a random topology.  In the greedy version
  we search for a local minimum (which can eventually be the global
  one), so it is important to start with a meaningful topology.  We
  tested the algorithm starting both with the FastME and the NJ
  reconstructed topologies.
  
\item[$2$] We sequentially consider all the internal edges of the
  present topology.  Each internal edge defines four subtrees (see
  fig.~\ref{fig:mcstep}), say $A,B,C,D$, rooted respectively on the
  four internal nodes $\alpha,\beta,\gamma,\delta$.  Referring to
  fig.~\ref{fig:mcstep}, let $((A,B),(C,D))$ being the initial
  configuration.  We randomly choose one of the two possible local
  rewirings: (i) swap the pair $B \leftrightarrow D $ getting the
  configuration $((A,D),(B,C))$, (ii) swap the pair $B \leftrightarrow
  C $ getting the configuration $((A,C),(B,D))$.  In the
  simulated-annealing version, the new configuration is accepted with
  a probability proportional to the statistical weight $e^{-\beta
    \mathrm{ sign}(\Delta E)}$ where $\beta$ is an inverse
  temperature-like parameter that is set by a simulated
  annealing-like~\cite{Kirkpatrick83} strategy and $\Delta E$ is the
  difference of the two configurations local costs: for case (i)
  $\Delta E = E_{((A,D),(B,C))} - E_{((A,B),(C,D))}$ while for case
  (ii) $\Delta E = E_{((A,C),(B,D))} - E_{((A,B),(C,D))}$.  In the
  greedy (or zero temperature) version, the new configuration is
  accepted if and only if $\Delta E < 0$.

\item[$3$] In the simulated-annealing version, we iterate point $2$ starting
  with $\beta=0$ and increasing it at a constant rate at each sweep
  (where we call {\em sweep} an update of all the internal edges) until
  convergence is reached, i.e., until the algorithm gets stuck in a
  fixed topology.  In the greedy version, we iterate
  point $2$ until the algorithm gets stuck in a fixed topology.

\end{itemize}

In order to define the configurational local cost, say
$E_{((A,B),(C,D))}$, we consider all the quartets $(ab:cd)$ such that
$a \in A$ ($a$ is a taxa of the subtree $A$) , $b \in B$, $c \in C$,
and $ d \in D$. For each quartet we define the {\em quartet
  frustration} as:

\begin{equation}
  \label{eq:fru}
  f_{(ab :cd)}=\max\left(0,\frac{D_1-\min(D_2,D_3)}{(D_1+\min(D_2,D_3))^k}\right),
\end{equation}

\noindent where $D_1,D_2,D_3$ are the sums of distances already
defined ($D_1=d(a,b)+d(c,d)$, $D_2=d(a,c)+d(b,d)$,
$D_3=d(a,d)+d(b,c)$). The normalization factor in the right-hand side
of Eq.~\ref{eq:fru}, as already pointed out in the introduction, gives
a smaller weight to longer distances, typically affected by noise and
recombination. The parameter fixing strategy for the exponent $k$ will
be discussed in the next subsection.

The cost $E_{((A,B),(C,D))}$ of the configuration $((A,B),(C,D))$ is
thus defined as the sum of the costs ({\em frustrations}) of all
the considered quartets, each weighted with a factor borrowed from the
Pauplin's formula:

\begin{equation}
\label{eq:quart-paup}
E_{((A,B),(C,D))}= \sum_{(ab :cd)} f_{(ab :cd)}
2^{-t(a,\alpha)-t(b,\beta)-t(c,\gamma)-t(d,\delta)},
\end{equation}

\noindent where $t(a,\alpha)$ is the topological distance between the
taxa $a$ and the internal node $\alpha$, and analogously for the other
taxa.

\subsection{Remarks on the cost definition}

When $k=0$ in Eq.~\ref{eq:fru} our procedure is equivalent to the
minimization of the Pauplin's length (the proof of this statement will
be discussed in the Appendix IA). On the other hand, if the
Pauplin weights $2^{-t(a,\alpha)}$, $2^{-t(b,\beta)}$,
$2^{-t(c,\gamma)}$, $2^{-t(d,\delta)}$ were absent, the difference in
local costs between two configurations would be equal to the variation
of a global cost, defined as $E= \sum_{(abcd)} f_{(ab :cd)}$. Here the
sum defining the cost, $\sum_{(abcd)}$, is running over {\em all} the
quartets of the tree and not only on the quartets compatible with the
sub-trees $A,B,C,D$. We will refer to our algorithm with this form of
the cost configuration as the {\em Normalized Quartets} method (NQ).

Conversely, when one takes the complete form of the local cost as
defined in Eq.~\ref{eq:quart-paup}, with $k>0$ in Eq.~\ref{eq:fru},
the local cost differences do not correspond to any global cost
difference (the proof of this statement will be discussed in Appendix
IB). It is however an open question whether a global functional can be
defined whose variation between each pair of configurations is
compatible with the sign of our local cost difference.

The complexity of a sweep of our algorithm (i.e., $N$ configurations
updates, where $N$ is the number of leaves in the tree), has a leading
term $O(N^4)$. The number of quartets to be considered when
  updating all the edges of the tree in case of a perfectly balanced
  tree reads:
\begin{equation}
{\cal N}= \frac{85}{5376}N^4-\frac{N^2}{3}+\frac{4}{7}N \,.
\end{equation}.
\noindent We show the results of numerical simulations for the running
time of the greedy version of our algorithm in Appendix IIB.
If considering the {\em Normalized Quartets} method, a na\"ive
minimization of the corresponding functional would lead to a
complexity of $O(N^5)$, while a Montecarlo sweep for a na\"ive
minimization of the Pauplin length is only $O(N^3)$. Despite the
$O(N^4)$ complexity of the SBiX algorithm, the greedy version has an
extremely low prefactor, making the algorithm suitable for trees with
a large number of taxa (see Appendix IIB).

\section{Results}

\subsection{Artificial phylogenies}

To test the performances of our algorithm, we consider artificially
generated phylogenies following one of the simplest evolutionary model
that takes into account both mutational events and horizontal
transfer. Each taxon is represented by a binary sequence of length
$l$. We start with one sequence, for instance the sequence with all
the bits equal to $0$. Then at each time step we perform the following
operations: (i) we randomly extract one of the already existing leaf
sequences, say $\bar{s}$; (ii) with probability $\tau$ a randomly
extracted portion of length $l/4$ of $\bar{s}$ is replaced with the
corresponding portion of another randomly chosen
sequence~\footnote{The choice of $l/4$ is arbitrary but does not bring
  loss of generality. Choosing randomly in the interval $[0,l/4]$ the
  length of the part of the sequence horizontally transferred does not
  alter the qualitative behaviour of the reconstructing algorithms
  (results reported in Appendix IIA). This last procedure is adopted
  in the $4$-state two parameters Kimura model (see below).}; (iii)
$\bar{s}$ generates two clones as descendants; (iv) each site of the
two new sequences is independently flipped with probability $m/l$,
where $m$ is extracted from an exponential distribution with average
$\mu$ (average number of mutations per sequence per time step). To
ensure that at least on site mutates at each branching event, we
randomly choose a site to mutate if no site mutated. We iterate this
procedure until the desired number of taxa is obtained.

We here consider as distance between two taxa the {\em correct hamming
  distance}~\cite{felsenstein2004}, defined as:

\begin{equation}
  d_{corr}= -\frac{1}{2} \ln(1- 2 h),
  \label{correct}
\end{equation}

\noindent where $h$ is the {\em hamming distance}, defined as the
fraction of sites in which the sequences differ\footnote{In all the
  results reported in this paper we let the algorithm infer the
  correct phylogeny by using the {\em correct hamming distance}
  $d_{corr}$. Even though the defined correction has its theoretical
  justification only in absence of horizontal gene transfer, we
  checked (data not shown) that using the {\em hamming distance} $h$
  all the considered algorithms show the same relative behavior as in
  the reported results, but the absolute performances are remarkably
  poorer.}.

Although the evolutionary model described above is a toy model for
describing evolution, it allows to control and to tune noise as well
as horizontal transfer events. We also test our algorithm on
phylogenies constructed following more realistic model of evolution,
such us the standard $4$-states two parameters Kimura
model~\cite{Kimura_1980}. In particular, we follow the same steps
described above for the $2$-state model, but we now consider sequences
of nucleotides, with an alphabet of $4$ letters, and different rates
of transitions ($\alpha$) and transversions ($2 \beta$). We consider
in this case as distance between two taxa the {\em correct hamming
  distance} for the Jukes-Cantor model~\cite{felsenstein2004} that is
the limit of the Kimura model when $\alpha=\beta$, which reads:

\begin{equation}
  d^{\mathrm{JC}}_{corr}= -\frac{3}{4} \ln(1- \frac{4}{3} h).
  \label{correct_K2P}
\end{equation}

\subsection{Robinson-Foulds measure}

In order to assess the performances of the different algorithms to
reconstruct the true phylogeny, we consider the standard
Robinson-Foulds measure (RF)~\cite{Robinson_Foulds_1981}, which counts
the number of bipartitions on which the inferred tree differs from the
true one. A bipartition is a split of the leaves in two sets realized
through a cut of a tree edge. We recall that it exists a one-to-one
correspondence between the bipartitions of the tree and the set of its
edges, so that each tree is uniquely characterized by the set of
bipartitions it induces. Note that, since we consider only binary
trees, the number of true positive bipartitions equals the number of
false positive bipartitions and both are equal to the RF measure.
 
\subsection{Competing algorithms}

In order to assess the performances of our algorithm, both in its
simulated annealing and greedy version, we compare it with the
Neighbor-Joining (NJ)~\cite{NJ} and the FastME~\cite{fastme_2002}
algorithms. In addition we implemented the Pauplin's length
minimization (from now onwards referred as PAUPLIN) by making use of
our SBiX algorithm in its form with $k=0$ (see above the section about
the algorithm's description). This in order to directly investigate
the effectiveness of a non-greedy minimization of the Pauplin's length
in reconstructing trees.  Finally we implemented the version of our
algorithm without the Pauplin's weights (NQ) (as discussed above).  We
also show a comparison with the performances of a state-of-the-art
character based algorithm, MrBayes~\cite{MrBayes}.

\begin{figure}[bp]
\begin{center}
\includegraphics[width=0.80\textwidth]{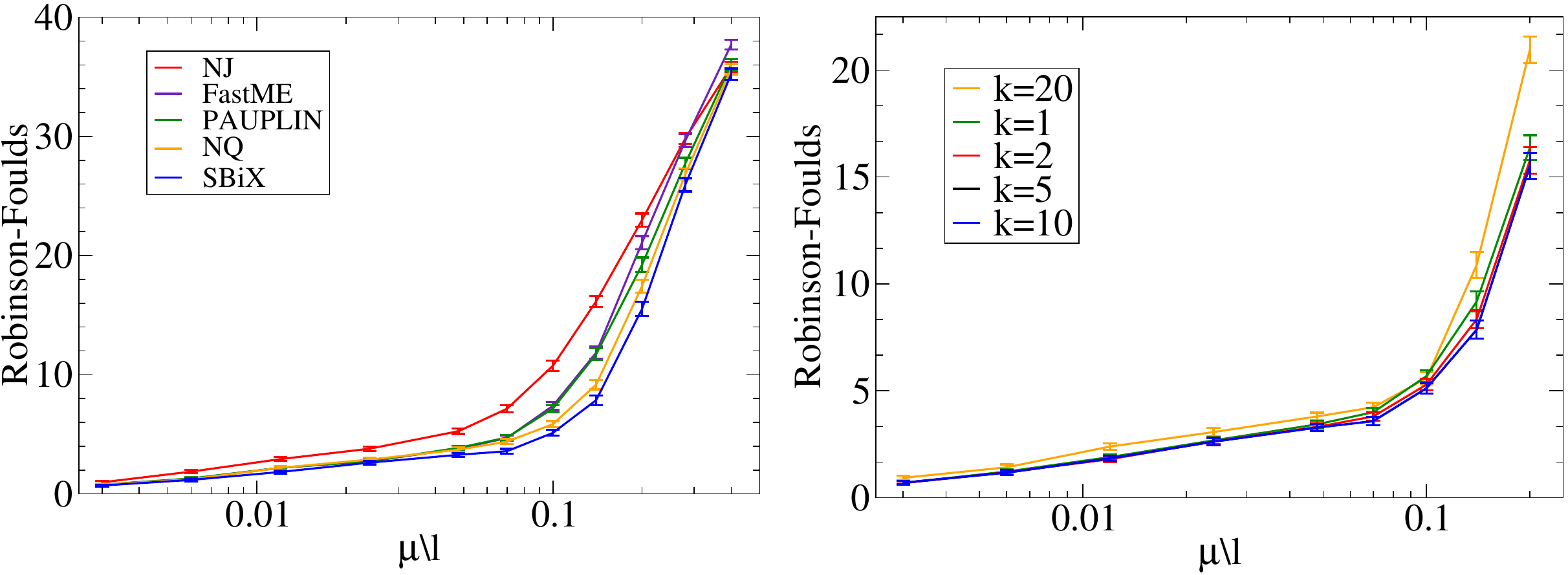}
\end{center}
 \caption{{\bf Performances comparison as a function of the mutation
     rate.} Left: Robinson-Foulds distance between the reconstructed
   and the true trees as a function of the mutation rate per site of
   the generative evolutionary model. The horizontal transfer rate
   $\tau$ is here kept $\tau=0$. We compare the performances of the
   SBiX algorithm with $k=5$ with: Neighbor-Joining (NJ), FastME, the
   Pauplin length minimization (PAUPLIN), and Normalized-Quartets
   (NQ). Right: dependence of the SBiX algorithm on the parameter $k$.
   The best performances are very stable in the range of $k$ between
   $5$ and $10$. In both figures results are averaged over $100$
   independent realizations for each reported mutation rate. The error
   bars are standard errors. All the trees generated have $N=60$
   leaves and the sequences have fixed length $l=1000$.}
\label{fig:rumandk}
\end{figure}

\begin{figure}[ht]
  \centerline{\includegraphics[width=0.80\textwidth]{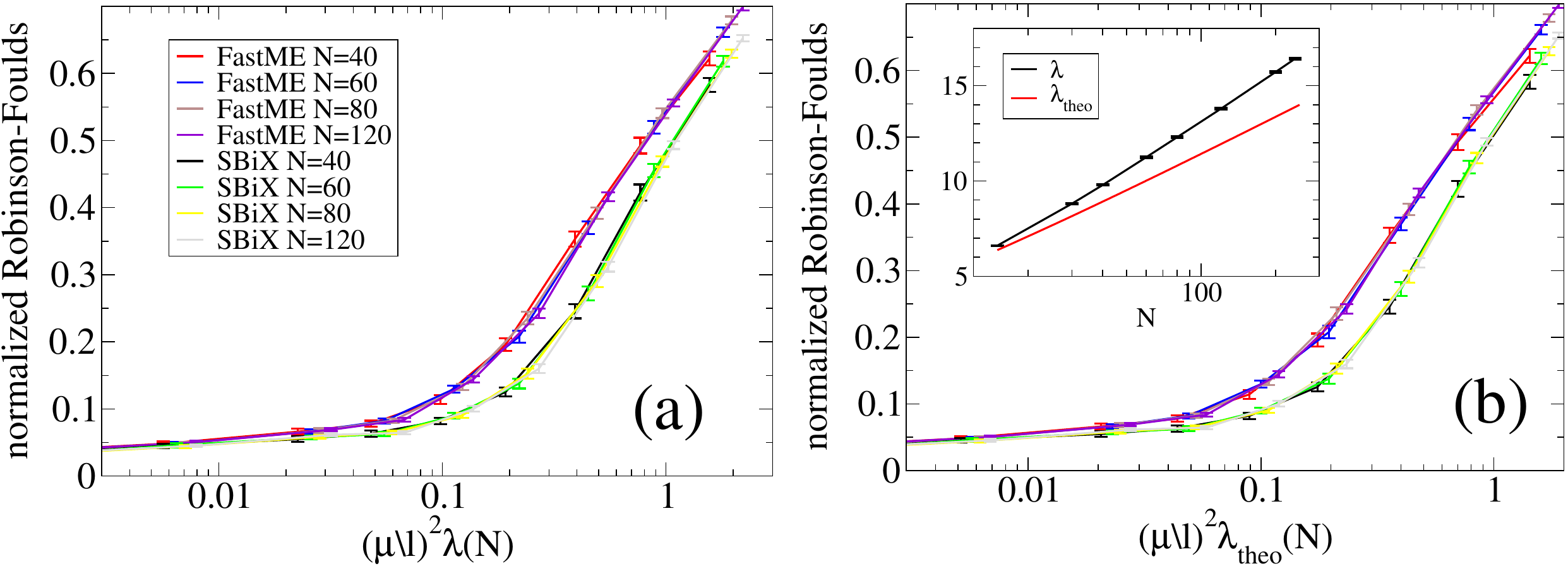}}
  \caption{{\bf System-size dependence.} Behavior of the normalized
    Robinson-Foulds distance for the SBiX algorithm and FastME for
    different system sizes, i.e., different values $N$ of the number
    of leaves. Here normalized means the Robinson-Foulds distance
    divided by its maximal value $N-3$. In all the cases curves for
    different values of $N$ collapse as a function of $\mu^2
    \lambda(N)$ (see text for details) where $\lambda(N)$ is the
    average distance between two leaves in a tree with $N$ leaves. In
    both the analysis the horizontal transfer rate $\tau$ is kept
    $\tau=0$. We use both the true value of $\lambda(N)$ in the
    simulated phylogenies {\bf (a)}, and the value analytically
    calculated in the case of perfectly balanced trees
    $\lambda_{\mathrm{theo}}(N)$ {\bf (b)}.  In the inset of {\bf (b)}
    we report the behavior of $\lambda$ and $\lambda_{\mathrm{theo}}$
    as a function of tree size $N$. The experimental values
    $\lambda(N)$ are systematically larger than
    $\lambda_{\mathrm{theo}}$ putting in evidence a slight deviation
    of the generated trees from a perfectly balance condition.}
\label{fig:collapse}
\end{figure}

\begin{figure}[ht]
\begin{center}
\includegraphics[width=0.80\textwidth]{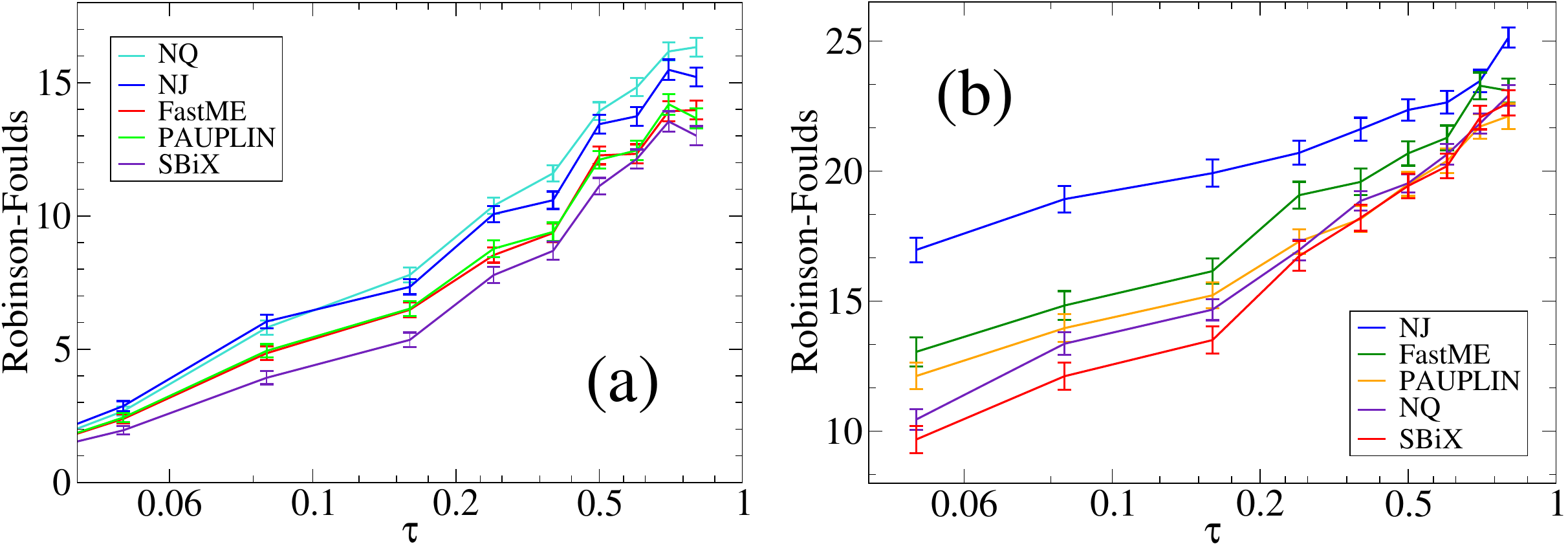}
\end{center}
  \caption{{\bf Performances comparison as a function of the
      horizontal transfer rate.} Robinson-Foulds measure for the
    reconstructed trees as a function of the horizontal transfer rate
    of the generative model. We compare the performances of our SBiX
    algorithm (with $k=5$) with those of the following ones:
    Neighbor-Joining (NJ), FastME, the Pauplin length minimization
    algorithm (PAUPLIN) and the Normalized-Quartet (NQ)
    algorithm. Results are averages over $100$ independent
    realizations for each reported horizontal transfer rate. The error bars are
    standard errors. All the trees generated have $N=60$ leaves. {\bf
      (a)}: mutation rate per site $\mu/l=0.003$ while the sequences
    have fixed length $l=10000$. {\bf (b)}: mutation rate per site of
    $\mu/l=0.14$ while the sequences have fixed length $l=1000$.}
\label{fig:res_hor}
\end{figure}

\subsection{Performances of the different algorithms}

In this section we compare the performances of all the considered
algorithms as a function of the mutation rate and of the horizontal
transfer rate in the underlying evolutionary process described in
the previous subsection.


In fig.~\ref{fig:rumandk} we show the Robinson-Foulds curves for
different algorithms (Left) as a function of the mutation rate for a
fixed tree size ($N=60$) and $k=5$. In the whole range of values of
the mutation rate all the versions of our algorithm (PAUPLIN, NQ and
SBiX) outperforms both NJ and FastME. In particular SBiX outperforms
all the other algorithms. Differences between the global minimization
of the Pauplin's length (PAUPLIN) and FastME arise for very high
mutation rates, where the global Pauplin's length minimization
outperforms FastME.  This is probably due to the fact that FastME is
time optimized and therefore less able of our Stochastic Local Search
scheme to find the global minimum of the functional for very high
mutation rates (for a discussion on the consistency of greedy local
moves based on the balanced minimum evolution principle
see~\cite{Bordewich_2009}).  In fig.~\ref{fig:rumandk} we report the
dependence of the SBiX performances (Right) on the value of the
parameter $k$. It is evident the existence of a range of values
between $k=5$ and $k=10$ where the algorithm features the best results
in a stable way. In the following, unless otherwise stated, we will
consider the $k=5$ case.

Up to this point we have characterized the performance of the
different algorithms for fixed value of the number of leaves, i.e.,
for a given system size. We are now interested in the robustness of
our results at different number $N$ of leaves of the tree. Defining
$\lambda(N)$ as the mean topological distance between any couple of
leaves, we empirically found that each algorithm can be characterized
by a reference curve obtained by plotting the normalized
Robinson-Foulds distance as a function of $\mu^2 \lambda(N)$. This
scaling can be understood by considering that the relevant quantity
for the tree reconstruction is not the bare mutation rate but the
amount of back mutation events, that can be estimated as $\mu^2
\lambda(N)$.

The scaling of the normalized Robinson-Foulds distance when
reconstructing trees of different sizes is shown in
fig.~\ref{fig:collapse}, where for the sake of clarity we only report
the curves for FastME and the SBiX algorithm.  Each of the two
algorithms is characterized by a different reference curve, and the
interesting point here is that the SBiX algorithm is systematically
better than FastME at all mutation rates and sizes. We use both the
measured value of $\lambda(N)$ in the simulated phylogenies
(fig.~\ref{fig:collapse}-(a)), and the value analytically calculated in
the case of perfectly balanced rooted trees
(fig.~\ref{fig:collapse}-(b)) that reads:
\begin{equation}
\lambda_{\mathrm{theo}}(N)=\frac{2 N (\log_2{N}+1)-4 N +2}{N-1}.
\end{equation}


We now consider the ability of the different algorithms in recovering
the correct tree in the presence of horizontal transfer events. The
Robinson-Foulds curves at a fixed tree size are shown as a function of
the probability $\tau$ for each sequence to receive a borrowing (with
the mechanism defined above). In fig.~\ref{fig:res_hor}-(a) results at low
mutation rate are reported, when deviation from additivity is almost
exclusively due to the horizontal transfer events. In
fig.~\ref{fig:res_hor}-(b), instead, results are reported at high
mutation rate, when both back mutations and the horizontal transfer
events are responsible for deviations from additivity.  For horizontal
transfer events co-occurring with a low mutation rate, our algorithm is
the most suitable to recover the correct tree, at each rate $\tau$.
The {\em Normalized Quartets} method, conversely, shows a performance
lower than that of NJ.  When a high mutation rate is considered
jointly with horizontal transfer events, our SBiX algorithm
significantly outperforms the others when the probability $\tau$ of
horizontal transfer is not too high, while in the high $\tau$ region
the performance of our algorithm becomes comparable to the
minimization of the Pauplin's length (PAUPLIN).

\begin{figure}[t]
\begin{center}
\includegraphics[width=0.40\textwidth]{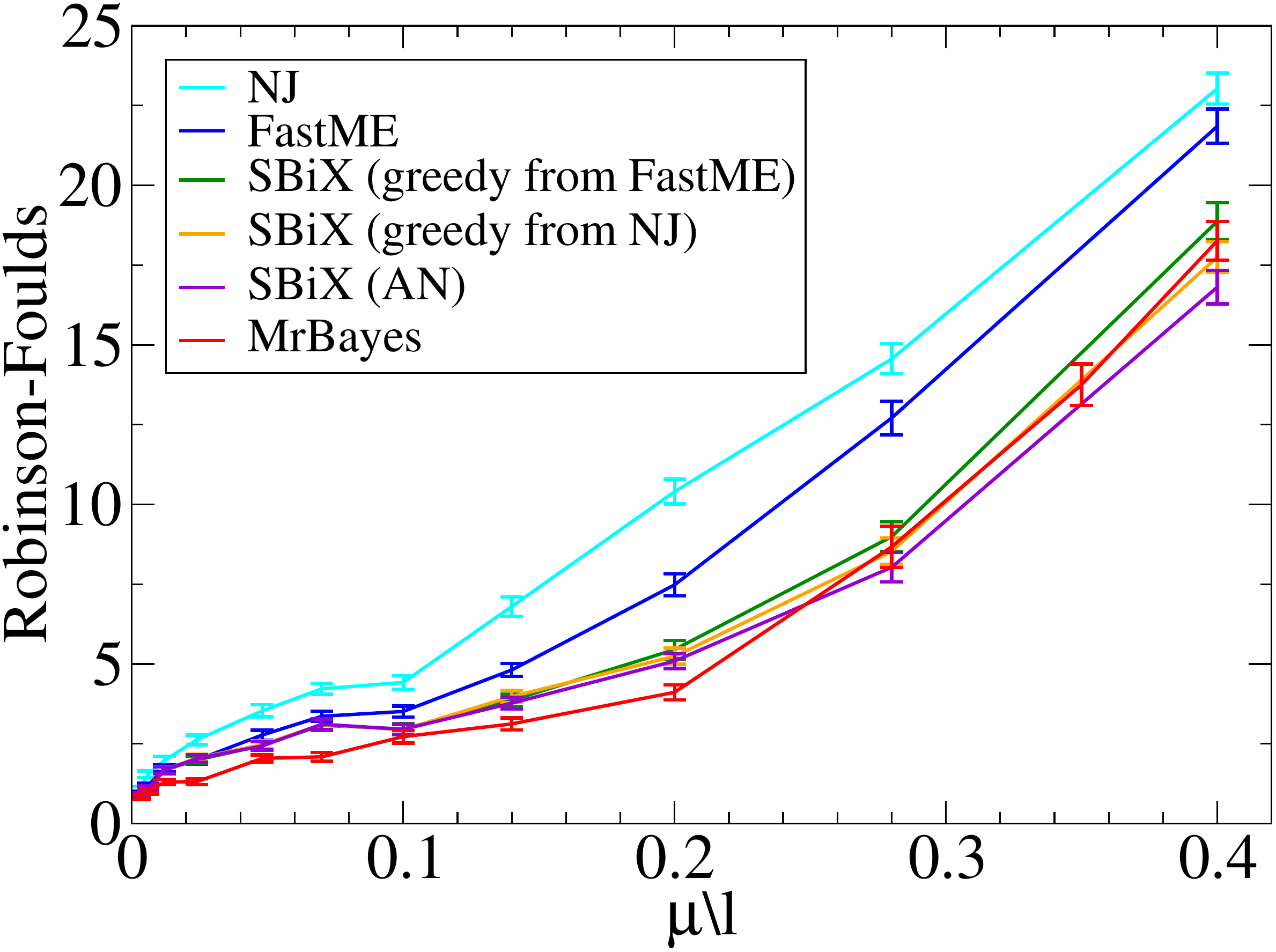}
\end{center}
\caption{{\bf Performances comparison as a function of the per-site
    mutations rate.}
 The evolutionary model used to generate
    phylogenies is the Kimura two parameters model with transition
    rate $\alpha=0.4$ and transversion rate $2 \beta=0.6$.
  Robinson-Foulds measure for the reconstructed trees as a function of
  the mutation rate per site. We compare the performances of our SBiX
  algorithm (with $k=5$) with those of the following ones:
  Neighbor-Joining (NJ), FastME, SBiX greedy from NJ, SBiX greedy from
  FastME, SBiX simulated annealing, and MrBayes. Results are averages
  over $100$ independent realizations for each reported mutation rate.
  The error bars are standard errors. All the trees generated have
  $N=60$ leaves, and the sequence $l=1000$.}
\label{fig:res_K2P}
\end{figure}

It is interesting to compare the performance of the different
algorithms in the case of a more realistic data generator. In
fig.~\ref{fig:res_K2P} we show the analogous of fig.~\ref{fig:rumandk}
for the two parameter Kimura model: we present the Robinson-Foulds
curves obtained for different algorithms (NJ, FastMe, SBiX greedy from
FastME, SBiX greedy from NJ, SBiX simulated annealing, and MrBayes)
for $N=60$ as a function of different per-site mutation rates $\mu/l$.
The details of the simulation for MrBayes are presented in Appendix
IIB, together with a comment on the computational complexity of the
whole simulation.

The first evidence is that SBiX, in its different flavors, clearly
outperforms the other two distance based algorithms (NJ, and
FastME). The improvement is even more evident compared with the binary
characters case displayed in fig~\ref{fig:rumandk}. The three variants
of SBiX perform similarly in the low mutation rate regime ($\mu/l
<0.2$), and the results show a moderate improvement of the simulated
annealing version only for high level of mutation rate, whereas the
difference between the two greedy versions of SBiX seem to be
statistically irrelevant in all the mutation rate interval analyzed.

The comparison with MrBayes is somehow surprising. After a very low
mutation rate regime ($\mu/l \leq 0.024$), where all algorithms show
analogous accuracies, we enter in an intermediate regime ($ 0.048 \leq
\mu/l \leq 0.24$) where MrBayes outperforms SBiX. One has to notice
that the improvement of SBiX vs. both NJ and FastME is more evident
compared with that of MrBayes vs. SBiX. In the high mutation rate
regime ($\mu/l > 0.24$) MrBayes shows results compatible with two
greedy versions of SBiX, while it is slightly outperformed by the
simulated annealing version of SBiX, only at the larger mutation rate
available ($\mu/l=0.4$).

One should be careful to draw conclusions about the performance of
MrBayes in the high mutation rate regime where the issue of reaching
the Monte Carlo Markov Chain steady state becomes really problematic.
We can not exclude that, upon doubling the simulation times in this
regime, one could improve MrBayes results. In this regime, as more
thoroughly discussed in Appendix IIB, the computational time for a
single sample is already of the order of five hours, whereas for both
greedy versions of SBiX is of the order of $10^{-2}$ seconds, and for
the simulated annealing version is around 5 minutes.

\section{Discussion and conclusions} 

In this paper we have introduced a new algorithmic scheme for
phylogeny reconstruction. Belonging to the family of Stochastic Local
Search algorithms, our scheme crucially exploits two known properties
of additive distance matrices, the {\em four-points condition} and the
so-called Pauplin's length. We proposed in particular a stochastic
scheme where the correct topology is inferred through a series of
swapping of the tree topology. When tested on artificially generated
phylogenies our algorithmic scheme significantly outperforms
state-of-art distance-based algorithms in cases of deviation from
additivity due to high rate of back mutations. A significant
improvement is also observed with respect to the state-of-art
algorithms in case of high rate of horizontal transfer.

Such good performances are due to the way we differentially weight the
different quartets contributions with a term inversely proportional to
their length, and thus to their probability to be affected by back
mutations. On the other hand further work is needed for a complete
theoretical understanding of the algorithm. In particular, despite
many attempts, we are, at present, unable to formulate the update
strategy in terms of a state functional. Beside the interest in
itself, this would open the way to analytic treatments as well as to
algorithmic optimization strategies possibly more efficient than the
Stochastic Local Search one.

As for the comparison of our algorithmic scheme with state-of-the art
algorithms it is fair to observe that SBiX features a definitely
larger computational complexity but, in practice, it is fast enough
for reconstructing phylogenies up to a few thousands of leaves.

Though SBiX outperforms all competitors also in presence of horizontal
transfers, the method is especially suited for dealing with
non-additivity originated by double mutations. The issue of horizontal
transfer is however central in many fields~\cite{Doolittle_2008}, and
we believe that formulating effective strategies to dealing with it,
considering both phylogenetic trees and networks, is an open challenge
for the next generation reconstruction algorithms and will be the aim
of further studies.

It is worth mentioning how the applicability of phylogenetic
algorithms has recently widened its scope. Many different fields have
arisen in the last few years where a correct reconstruction of
phylogenetic trees may reveal underlying relevant dynamical processes.
For instance phylodynamics is a new field at the crossroad of
immunodynamics, epidemiology and evolutionary biology, that explores
the diversity of epidemiological and phylogenetic patterns observed in
RNA viruses of vertebrates~\cite{grenfell_2004}; phylogeography is
the study of the historical processes that may be responsible for the
contemporary geographic distributions of individuals as well as of
languages or viruses~\cite{avise_2000}. In all these cases a strong
effort is being devoted to the collection of comprehensive datasets
and efficient and reliable algorithms are needed especially when
deviations from perfect phylogenies become relevant.

\section{Appendix I}
\label{app_I}

\subsection{The equivalence with the Pauplin length minimization}

We show here that our SBiX algorithm minimizes the Pauplin's length
when $k=0$ in the Eq.~\ref{eq:fru}. In order to see this, we
explicitly calculate the cost difference between, say, the
configurations $((A,B),(C,D))$ and $((A,C),(B,D))$. We note that the
sum on the considered quartets can be divided in three parts, in which
one of the three distances $D_1$, $D_2$ and $D_3$ is respectively
minimal (where, as already defined in the text, $D_1=d(a,b)+d(c,d)$,
$D_2=d(a,c)+d(b,d)$ and $D_3=d(a,d)+d(b,c)$). After a little algebra
one gets:

\begin{equation}
  \Delta E =  \sum  \left( D_2-D_1 \right)  2^{-t(a,\alpha)-t(b,\beta)-t(c,\gamma)-t(d,\delta)},
\end{equation}

\noindent where the sum is again over the $a \in A, b \in B, c \in C,
d \in D$ and $\Delta E \equiv E_{((A,C),(B,D))}-E_{((A,B),(C,D))}$.
Making use of the relation:

\begin{equation}
  \sum_i 2^{-t(i,r)}=1,
\end{equation}

\noindent and of the equivalences:
$2^{-t(a,b)}=2^{-t(a,\alpha)-t(b,\beta)}/2$ in the configuration
$((A,B),(C,D))$ and $2^{-t(a,b)}=2^{-t(a,\alpha)-t(b,\beta)}/4$ in the
configuration $((A,C),(B,D))$ (and the analogous relations for the
other pairs of taxa), it is easy to prove that it holds:

\begin{equation}
  \Delta E= 4 \Delta L_P,
\end{equation}

\noindent where $L_P$ is the Pauplin's length and $\Delta L_P \equiv
L_{P,((A,C),(B,D))}-L_{P,((A,B),(C,D))}$ is the difference of the
Pauplin's length between the two configurations.

\subsection{Locality of the SBiX configuration cost}

We give here an argument to prove that differences in the local cost
of our SBiX method cannot be written as differences of a functional on
the whole tree. If this was the case, a functional could be defined as
$F(x)=F(x_0)+ \sum\Delta E_i$, where $F(x_0)$ is the value taken by
the functional in a reference configuration $x_0$, and $\Delta E_i$
are the cost differences along a path from $x_0$ to $x$. Moving in the
space of tree's topologies, we should obtain the same value of $F$
each time we visit the same topology, i.e., the difference of cost
between two states does not depend on the path. This is not the case,
as we explicitly checked, when the cost is defined as in
Eq.~\ref{eq:quart-paup} and $k \neq 0$.

\section{Appendix II}
\label{app_II}

\subsection{Discussion on the horizontal gene transfer modeling}

\begin{figure}[htb]
\centerline{\includegraphics[width=0.40\textwidth]{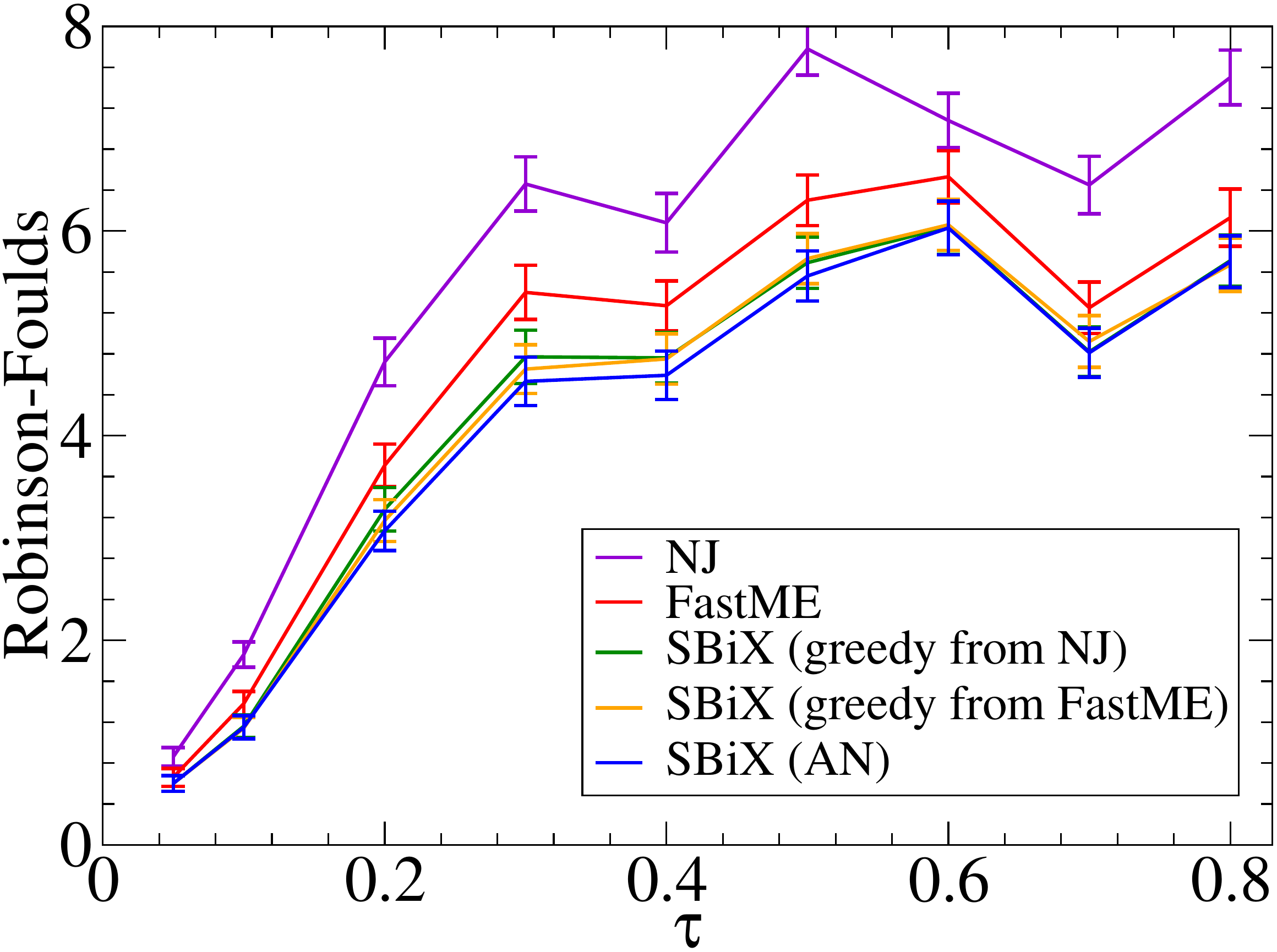}}
\caption{{\bf Performances comparison as a function of the horizontal
    transfer rate.} Robinson-Foulds measure for the reconstructed
  trees as a function of the horizontal transfer rate of the
  generative model. We compare the performances of our SBiX algorithm
  (with $k=5$) with those of the following ones: Neighbor-Joining
  (NJ), FastME, the Pauplin length minimization algorithm (PAUPLIN)
  and the Normalized-Quartet (NQ) algorithm. Results are averages over
  $100$ independent realizations for each reported horizontal transfer
  rate The error bars are standard errors. All the trees generated
  have $N=60$ leaves, the mutation rate per site is kept fixed to
  $\mu/l=0.003$ while the sequences have fixed length $l=10000$.}
\label{fig:hor_nofix}
\end{figure}

In order to assure that the procedure of keeping the length of the
horizontal transfer fixed to $l/4$, as for the results shown in the
main text, does not alter the qualitative behaviour of the compared
performances, we here show results where this restriction is relaxed.
The Figure~\ref{fig:hor_nofix} is the analogous of Figure 5 (a) in the
main text, with the difference that here the proportion of horizontal
gene transfer is allowed to randomly vary in the interval [0,1/4].

\subsection{Discussion on the algorithmic computational complexity}

\subsubsection{Generalities}

Given the encouraging performance of the SBiX compared to other
distance based algorithm we have run a detailed comparison with
character based methods such as MrBayes \cite{MrBayes}. For all our
comparison we used {\tt MrBayes v3.1.2} on a standard linux platform,
Intel Core(TM)2 CPU $@$ $2.40 GHz$ and a cache
size of $4096 KB$. In the simulation we kept the default setting of
the program ( {\tt Nruns = 2, Datatype = DNA, Nucmodel = 4by4, Nst =
  1, Covarion = No, \# States = 4, Rates = Equal} ) and we did not
specify any details of the model generator. The only parameters we
tuned were: (i) the number {\tt Ngen} of tree generated by the Monte
Carlo Markov Chain (MCMC), (ii) the number {\tt Burnin} of tree
configurations discarded in the analysis. Parameter fixing depends
critically on data, and in particular, the more the generative model
departs from additivity, the longer the MCMC must be run in order to
achieve convergence and obtaining a fair sampling of the likelihood
landscape of the problem. A useful proxy for MCMC convergence is the
average standard deviation of the partition split frequencies in the
{\tt Nrun} (in our case {\tt Nrun=2}) MCMCs. One expects that this
quantity, after a transient, approaches to zero for a sufficiently
long number of iterations, reflecting the fact that the two Markov
Chains become increasingly similar, sampling the same subspace of the
problem.

\subsubsection{Computational complexity}

Assessing precisely the computational complexity of MrBayes goes
definitely beyond the scope of our work. We performed however a
detailed analysis on a particular case for giving a fair comparison of
the computational cost of the two strategies.

As a benchmark we used the two parameters Kimura generator (with
transition rate $\alpha=0.4$ and transversion rate $2 \beta=0.6$)
discussed in the paper. We generated 100 realization of phylogenies,
with sequence length $l=1000$, with a per-site mutation rate $\mu/l =
0.1$, at different values of the number of leaves ranging from $N=15$
to $N=480$.

As a lower bound on the computational cost of MrBayes we decided
to stop the simulation as soon as the average standard deviation of
the partition split frequencies became lower than $0.05$. We intend
this as a lower bound on the true computational cost since this
can be considered as a good estimation only of the length of the
transient regime.  To ensure a fair {\em equilibrium sampling} one
should check that the 2 Markov Chains maintains split values lower
than our threshold for a number of generations large enough. For each
of our samples we recorded the running time in seconds and we averaged
over the 100 sample realizations.

\begin{table}[htb]
\begin{tabular}{|*{8}{c|}}
  \hline
  $N$ & $\overline{t}_{\mathrm{mb}}$ & var$(t)_{\mathrm{mb}}$& samples
  (mb) &  $\overline{t}_{\mathrm{SBiX}}$ & var$(t)_{\mathrm{SBiX}}$  &
  samples (SBiX) & ratio $\overline{t}_{\mathrm{mb}} /\overline{t}_{\mathrm{SBiX}}$  \\ 
  \hline
  &&& &\\[-1.3em]
  \hline
  15  & 11    & 1   & 50 & NA    & NA   & NA & NA\\[-0.4em]
  30  & 60    & 3   & 50 & 0.004 & 0.04 & 100 & $1.5 \times 10^4$\\[-0.4em]
  60  & 297   & 28  & 50 & 0.04  & 0.01 & 100 & $7.4 \times 10^3$\\[-0.4em]
  120 & 2277  & 258 & 50 & 0.64  & 0.02 & 100 & $3.5 \times 10^3$ \\[-0.4em]
  240 & 17106 & 692 & 10 & 10.72 & 0.03 & 100 &$1.5 \times 10^3$ \\[-0.4em]
  480 & NA    & NA  & NA &200.67 & 0.03 & 100 & NA\\[-0.4em]
  960 & NA    & NA  & NA &5096.7 & 0.3 & 100 & NA\\
  \hline
\end{tabular}
\protect\caption{In this table we display the number of leaves $N$,
  the average running time in seconds $\overline{t}$, its variance,
  and the number of samples, both for MrBayes and SBiX algorithms. 
  The last column reports the ratio between the average running times
  of  MrBayes and SBiX.}
\label{tab:times}
\end{table}

On the same dataset we run the greedy version of SBiX starting from
the configuration predicted by Neighbor Joining, and recorded the
running time in seconds. In Table~\ref{tab:times} we displays the
comparison of the two algorithms. We did not display results for SBiX
at $N=15$, the execution time being so tiny (order of milliseconds)
that the measure would be strongly unreliable and basically dominated
by input-output operating system tasks. We also do not present results
for MrBayes at $N=480$, since each run would require a time larger
than 32 hours.

\begin{figure}[htb]
\centerline{\includegraphics[width=0.55\textwidth]{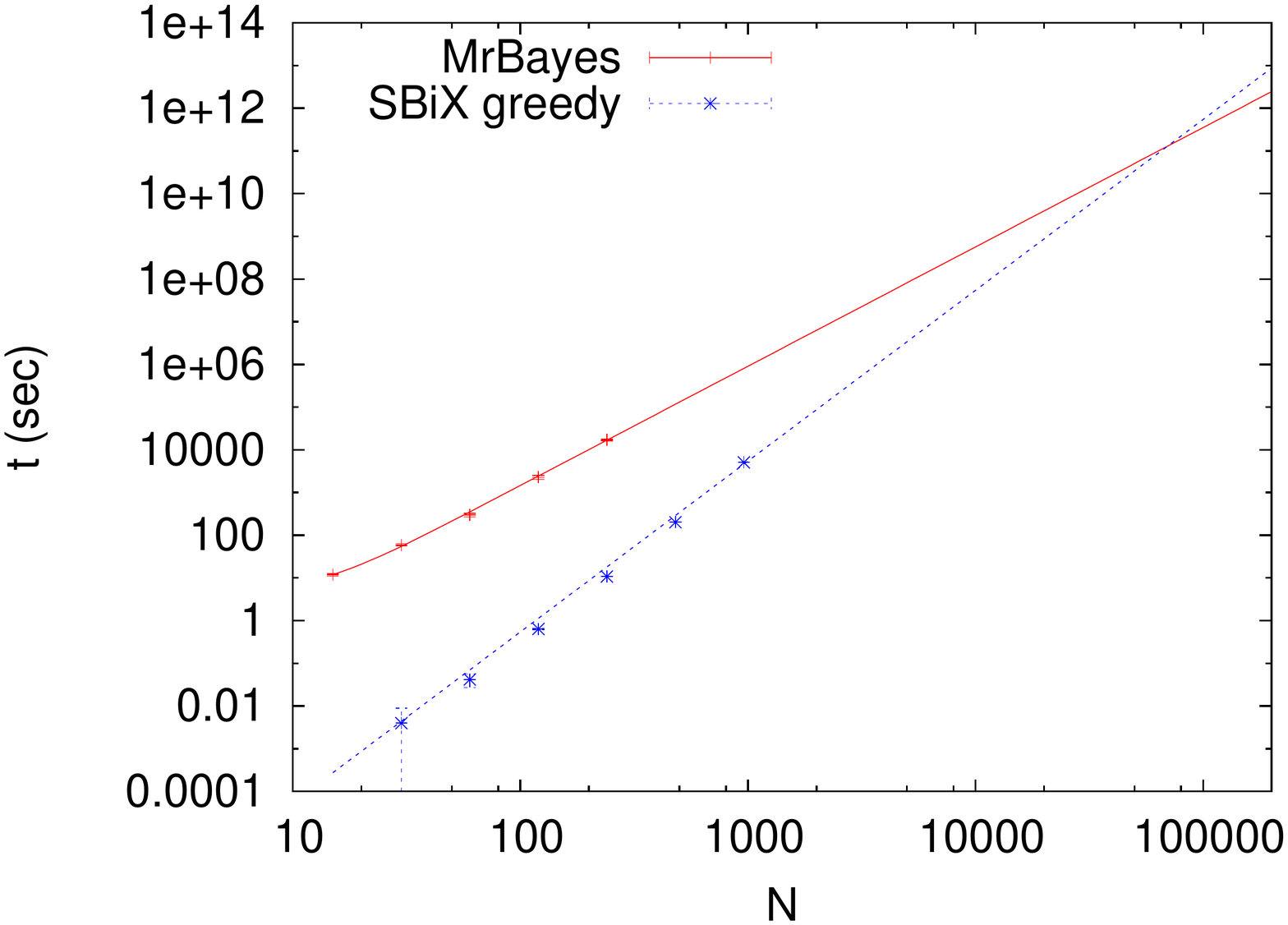}}
\caption{Plot of the average time of computation per sample (in
  seconds) as a function of the number of leaves of the phylogeny. All
points represent an average over the respective number of samples as
indicated in Table~\ref{tab:times}}
\label{fig:sbix_vs_mb}
\end{figure}

In Figure~\ref{fig:sbix_vs_mb} we plot the average running time per
sample as a function of $N$ for both algorithm. We fitted MrBayes
result with a simple power-law $ a + b x^c $ obtaining as best fit
$a=4(1)$ , $b=0.003(1)$, $c=2.8(1)$. The SBiX curve can be fitted with
a simple fourth power behavior $ A x ^4 $, and the best fit value is
$A = 5.4(4) \cdot 10^{-9}$. Such a small prefactor allows SBiX to be
faster than MrBayes to $N \sim 10^5$, despite the larger exponent (4
vs. 2.9) . Note that: (i) the running times at the crossing point are
of the order of the thousand of years, and (ii) hardly a $10^5$ leaves
genealogy could be of any interest.

\subsubsection{Computational cost as a function of the mutation rate}

In the previous section we outlined the analysis of the computational
complexity as a function of the number of taxa in the genealogy. In
the paper we presented the accuracy of the different algorithms fixing
the number of taxa ($N=60$) and varying the mutation rate per site
$\mu/l$. While the running time of other algorithms (NJ, FastME, SBiX
greedy) do not depend on the mutation rate, the running time required
by MrBayes depends critically on it. In order to obtain a fair
sampling of the space of trees we have run a number of preliminary
simulations in order to have a quantitative estimate of the
convergence time of the algorithm (the so-called {\em burn-in}
time). Then we considered a measuring time either $50000$ or $100000$
Monte Carlo steps depending on the mutation rate (and on the burn-in
time).  In Table~\ref{tab:mc} we display the burn-in and measure times
for the different values of the mutation rate. For each of the point
we analyzed 100 different samples. The convergence time in the high
mutation rate regime ($\mu/l >= 0.2$) might be still too small, and
for some of the samples the MCMC might not have reached
convergence. In order to guarantee convergence for all samples one
should double the burn-in time, requiring a time too long for our
computational resources.

\begin{table}[!htb]
\begin{tabular}{|*{4}{c|}}
\hline
$\mu/l$ & burn-in & measure & sec. \\
\hline
&&&\\[-1.3em]
\hline
0.003 & 50000  & 50000  & 1300  \\[-0.40em]
0.006 & 50000  & 50000  & 1300  \\[-0.40em]
0.012 & 50000  & 50000  & 1300  \\[-0.40em]
0.024 & 50000  & 50000  & 1300  \\[-0.40em]
0.048 & 50000  & 50000  & 1300  \\[-0.40em]
0.070 & 50000  & 50000  & 1300  \\[-0.40em]
0.100 & 100000 & 100000 & 2600  \\[-0.40em]
0.140 & 100000 & 100000 & 2600  \\[-0.40em]
0.200 & 300000 & 100000 & 5200  \\[-0.40em]
0.280 & 300000 & 100000 & 5200  \\[-0.40em]
0.350 & 500000 & 100000 & 7800  \\[-0.40em]
0.400 & 700000 & 100000 & 12400 \\
\hline
\end{tabular}
\protect\caption{In this table we display, for the different value of
  the per-site mutation rate $\mu/l$, the burn-in and the measure time in
  elementary Monte Carlo steps unit, together with an average estimate
  of running time in seconds. }
\label{tab:mc}
\end{table}

\section{Acknowledgments}
The authors wish to thank Simone Pompei, Francesca Colaiori, Ciro
Cattuto and Dami\'an H. Zanette for many interesting discussions and
suggestions. This research has been partly supported by the TAGora and
ATACD projects funded by the European Commission under the contracts
IST-34721 and 043415.


\end{document}